# Non-Invasive Photodelamination of van der Waals Semiconductors for High-Performance Electronics

*Ning Xu†, Xudong Pei†, Lipeng Qiu, Li Zhan, Peng Wang, Yi Shi*, Songlin Li**

† These authors contributed equally to this work.

E-mails: sli@nju.edu.cn and yshi@nju.edu.cn

N. Xu, L. Qiu, L. Zhan, Prof. Y. Shi, Prof. S. Li

School of Electronic Science and Engineering, National Laboratory of Solid-State Microstructures and Collaborative Innovation Center of Advanced Microstructures, Nanjing University, Nanjing 210023, China

X. Pei

College of Engineering and Applied Sciences and Jiangsu Key Laboratory of Artificial Functional Materials, Nanjing University, Nanjing 210023, China

Prof. P. Wang

Department of Physics, University of Warwick, CV4 7AL Coventry, UK

**ABSTRACT:** Atomically thin two-dimensional (2D) van der Waals semiconductors are promising candidate materials for post-silicon electronics. However, it remains challenging to attain completely uniform monolayer semiconductor wafers free of over-grown islands. Here, we report the observation of the energy funneling effect and ambient photodelamination phenomenon in inhomogeneous few-layer $WS_2$ flakes under low illumination fluencies down to several nW/μm$^2$ and its potential as a non-invasive post-etching strategy for selectively stripping the local excessive overlying islands. Photoluminescent tracking on the photoetching traces reveals relatively fast etching rates around 0.3−0.8 μm/min at varied temperatures and an activation energy of 1.7 eV. By using crystallographic and electronic characterization, we also confirm the non-invasive nature of the low-power photodelamination and the highly preserved lattice quality in the as-etched monolayer products, featuring a comparable average density of atomic defects (~$4.2\times10^{13}$ cm$^{-2}$) to pristine flakes and a high electron mobility up to 80 cm$^2$ V$^{-1}$ s$^{-1}$ at room temperature. This approach opens a non-invasive photoetching route for thickness uniformity management in 2D van der Waals semiconductor wafers for electronic applications.

## 1. Introduction

The success of modern semiconductor industry benefits vastly from the ever-increasing processing and etching capability over semiconductor channels for miniaturized device units. As the microelectronics based on bulk silicon approaches its limit, atomically thin two-dimensional (2D) van der Waals crystals will be required in near future, owing to their ideal dimensionality and self-saturated surface atomic structures.[1–3] Although impressive progress has been recently achieved in direct bottom-up synthesis of highly oriented 2D semiconductor wafers,[4–7] the yield of purely uniform wafers remains low and, in particular, the rate of thickness uniformity over the entire wafers, ~99% in reported results, is yet to be improved to meet the stringent requirement for electronic applications. In this context, one of the grand challenges for synthesized 2D semiconductor wafers can be attributed to suppression of the non-uniform nucleation and removal of the local excessive overlying islands. Hence, a highly precise post-etching strategy that enables selectively stripping the over-grown overlying layer(s) (i.e., the extra 2nd or/and 3rd top layers) from the desired monolayer wafers represents a useful scheme for thickness uniformity optimization.

Top-down thinning schemes with various exciters, including laser,[8–12] plasma,[13–15] and oxidants,[16,17] have been proposed, which are proven effective in reducing the thickness of van der Waals semiconductors but, in general, suffer from the issue of either lack of etching controllability or invasivity to the underlying monolayers to be preserved, because most of the schemes involve aggressive physical or chemical reactions that are prone to cause unselective over-etching.[8–12] It is well recognized that the electronic performance (e.g., charge mobility) of 2D semiconductors is highly sensitive to lattice defects or interfacial Coulomb impurities[18,19] and, thus, the traits of high processing controllability and non-invasivity represent vital but challenging prerequisites for thickness uniformity management in 2D semiconductors from the point view of practical applications. Very recently, we have exploited a realistically non-invasive etching scheme by employing low-temperature alloying and post-etching.[20] However, the processing convenience of this scheme still falls short of expectations.

In this work, we report the observation of an emerging photodelamination phenomenon in few-layer van der Waals semiconductors and its potential as a straightforward atomic-layer etching strategy under low-power light illumination. We find that few-layer $WS_2$ can be directly thinned down to monolayers upon excitation down to the level of nW/μm$^2$, where the overlying top layers are photooxidized into amorphous aggregates and can be simply stripped by rinse. The consequence of thickness reduction after photodelamination is jointly verified by multiple spectral methods such as photoluminescence (PL) and Raman characterization. The etching rates reach as high as 0.3−0.8 μm/min at varied temperatures, featuring an activation energy of 1.7 eV in the photochemical reaction. Importantly, we identify the merit of non-invasivity in this less aggressive photodelamination phenomenon by coincident crystallographic and electronic characterization. The density of atomic vacancies in the as-etched monolayer products is estimated to be as low as ~$4.2\times10^{13}$ cm$^{-2}$ and the intrinsic electron mobility reaches up to 80 cm$^2$V$^{-1}$s$^{-1}$ at room temperature; both values are comparable to the pristine counterparts. This finding represents a non-invasive but convenient uniformity management strategy in 2D van der Waals semiconductors for high-performance electronics.

## 2. Results and Discussion

### 2.1 Photodelamination phenomenon

Prior to introducing the intriguing photodelamination phenomenon, we first clarify its differences in excitation power and reaction mechanism with the conventional high-power laser thinning,[8–12] in spite of their similarity in etching exciters (**Figure 1**a). The conventional laser thinning relies mostly on the effects of instantaneous photoheating and thermal sublimation of the local regions of 2D materials, in which the use of high laser fluencies is a typical character and the local sample temperature can reach ~1698 K at 90 mW/μm$^2$ (inset of Figure 1a).[10] By contrast, the variation of sample temperature in photodelamination is negligible (e.g., ~0.08 K under 60 nW/μm$^2$), in which low fluencies are often adopted and, essentially, the photochemical reaction between the material edges and the absorbed aqueous oxygen from ambient surroundings is responsible for the controllable oxidation of the overlying layers of $WS_2$ (Figure 1a).[21,22] Hence, the photodelamination occurs only from the flake edges (See Movie S1, Supporting Information) where the active reaction sites are located due to the presence of dangling atomic bonds, while the photothermal thinning can be initiated anywhere in a fashion as direct writing.[8–12]

To present the photodelamination behavior vividly, Figures 1b and 1c show contrastive optical images and schematic atomic structures before and after light illumination (λ = 455 nm, 24˚C, relative humidity ~60%) for a mechanically exfoliated $WS_2$ flake with varied local thicknesses: monolayer (1L), bilayer (2L) and bulk. Here the exfoliation medium poly(dimethylsiloxane) (PDMS) is directly used as the supporting substrate for convenience. In the beginning, the $WS_2$ flake contains mixed pristine 1L and 2L areas (Figure 1b). Note that the thickness information can be easily obtained from the brightness or photoluminescent (PL) contrast (insets of Figure 1b), because the 1L area (direct bandgap) generates much stronger PL emission than the thick areas (indirect bandgap).[23,24] After appropriate illumination, the 1L area expands and covers the 2L area, that is, the original 2L area is converted into 1L entirely. We term this behavior as photodelamination, because the underlying layer in the 2L area is well preserved and only the overlying layer is consumed. From the PL image shown in the inset of Figure 1c, one can find that the PL intensity of the newly converted 1L area is almost identical to the pristine 1L area on its left, indicating the high crystalline quality of the converted 1L area.

Raman and PL spectra were also used to verify the consequence of the delamination behavior by comparing the corresponding spectral characteristics for the 2L area before and after photodelamination. Figure 1d shows the contrastive Raman spectra to demonstrate the effect of photodelamination, where the intervals between the two $A_{1g}$ and $E_{2g}$ Raman peaks are 62.3 and 60.5 cm$^{-1}$ before and after photodelamination, respectively. The change in the interval of Raman peaks is consistent with the effect of thickness tailoring from 2L to 1L.[25] Likewise, contrastive PL spectra (Figure 1e) also support the consequence of thickness reduction. For the pristine 2L, the PL intensity is negligible as compared with the pristine 1L, because of its indirect bandgap.[23,24] However, after photodelamination its PL intensity soars dramatically and becomes comparable to the pristine 1L, suggesting its conversion to a direct bandgap semiconductor due to the reduction in effective thickness[23] without sacrificing the optical properties of the as-thinned bottom layer. These spectral characteristics strongly corroborate the photodelamination in nature, as well as its high etching accuracy down to atomic scale.

Activation energy responsible for the photodelamination process was also estimated through variable temperature experiment. Real-time PL tracking on photoetching traces (Movie S1, Supporting Information) reveals relatively fast etching rates around 0.3−0.8 μm/min at varied temperatures (Figure 1f). According to the classical Arrhenius rule, the relation between reaction rate ($r$) and activation barrier ($E_a$) can be described as $r \propto e^{-\frac{E_a}{kT}}$, where $k$ is the Boltzmann constant, and $T$ is the absolute temperature. Figure 1f shows the Arrhenius plot for $r$ versus 1000/$T$. From the slope we extracted $E_a$ = 1.7 ± 0.4 eV.

## 2.2 Reaction mechanism

To look insight into the reaction mechanism, we employed micro-zone elemental analysis techniques to characterize the reaction products. **Figure 2**a shows the micro-zone scanning transmission electron microscopy (STEM) image for a fully reacted 1L $WS_2$ flake, where tiny aggregates with lateral sizes of few nanometers are randomly distributed. The circular STEM diffraction patterns (inset of Figure 2a) indicate that the reaction products are of fully amorphous state and the pristine lattice crystallinity is completely destroyed. Elemental mappings were further used to assess its chemical compositions, which suggests the presence of excessive oxygen (O) element, in addition to the tungsten (W) and sulfur (S) elements that exist in pristine $WS_2$. Notably, the intensity of the element oxygen is much stronger than that of sulfur, indicating that the photodelamination phenomenon is essentially a surfacial photooxidation process, in which ambient oxygen is involved actively.[21]

To analyze the valence state of tungsten in the reaction products, X-ray photoelectron spectroscopy (XPS) characterization was performed on a light illuminated thick flake. Note that the reason to use large-area but thick, rather than small and thin, flake is just to fit the size (~100 μm) of the x-ray source in XPS. Figure 2b plots the spectrum for the tungsten 4$f$ core level spanning from 31 to 40 eV, where four characteristic peaks were detected. The peaks at 33.0 and 35.2 eV are assigned to the excitations from $W^{4+}$ atoms, arising likely from the underlying unreacted $WS_2$ layers, while the high-lying peaks at 36.2 and 38.1 eV are assigned to the oxidized $W^{6+}$ atoms, stemming from the reacted oxysulfide or oxides.[26,27] This observation further corroborates the ambient photooxidation mechanism behind the photodelamination phenomenon.

## 2.3 Removal of reaction products

Next, we explore the feasibility of exploiting the nontrivial photodelamination phenomenon as a selective atomic-layer etching strategy, in which the reaction products are expected to be readily stripped after etching. For instance, the gaseous reaction products are normally pumped out in the conventional reactive ion etching; the solid products are evaporated under instantaneously high temperatures in the photothermal thinning.[8–12] However, in the photodelamination process, the reaction products develop into unwanted solid residues staying on the surfaces, constituting a possible issue for device fabrication.

To address this issue, we further employed atomic force microscopy (AFM) to analyze the adhesive strength of the amorphous reaction products to the underlying preserved $WS_2$ layers. Figure 2c shows the surface morphology for another partially photodelaminated flake, where the reaction was intentionally terminated in the middle without consuming all the 2L area and thus, the image captures three representative local areas: pristine 1L, photoetched/thinned 1L (uneven morphology), and pristine 2L. In the photoetched area, randomly distributed aggregates up to a micron scale are seen, which can be attributed to the delaminated amorphous reaction products that accumulate due to fluctuations of local strain and temperature in the light illumination. We found that these reaction products are loosely attached to the photoetched region such that they can be easily stripped by moderate rinse in deionized water, because all the delaminated aggregates disappear from the photoetched region after rinse, as shown in the contrastive AFM image (Figure 2d). A clean surface can be obtained hereafter.

Apart from the residue stripping, we also checked the etching precision of the photodelamination. Figure 2e shows the surface morphology near the etching step for the partially photoetched sample after stripping the surficial reaction residues. A height difference of 0.68 nm is recorded, which corresponds to the thickness of a $WS_2$ monolayer, verifying the atomic etching precision. Such high etching precision is also proven by the cross-sectional STEM image given in Figure 2f. Both results indicate that the photodelamination and subsequent rinse enable a clean and precise tailoring of the van der Waal semiconductors.

### 2.4 Energy funneling effect and reaction selectivity

In addition to the 2L samples, we find that the photodelamination also occurs in the thicker samplers with 3L areas, as shown in Figure S1 (Supporting Information). We then elaborate on the mechanism of reaction selectivity behind the photodelamination phenomenon, which can be readily attributed to the presence of ordered band alignment and resultant energy funneling effect in the thickness-varied inhomogeneous semiconductors due to quantum confinement.[23,28] **Figure 3** shows the schematic diagram for an ambiently placed $WS_2$ flake with mixed 1−3L local areas (Figure 3a) and corresponding energy levels for the local areas and aqueous oxygen absorbed at the $WS_2$ surfaces (Figure 3b). In the ambient surroundings filled with humidity and oxygen, the etching selectivity originates mainly from two factors: 1) Electron funneling transfer and accumulation in thick areas, and 2) Exact locations of the reaction sites.

It was reported that electrochemical oxidation is the responsible mechanism for the degradation of nanoscale semiconductors in ambient surroundings,[21] where the reaction occurs between semiconductors and aqueous oxygen absorbed on surfaces. As depicted in Figure 3b, incident light illumination can facilitate the reaction by exciting electrons from the valence ($E_V$) to conduction band ($E_C$) in semiconductors to promote the generation of highly active superoxide ions $O^{2-}$ in aqueous oxygen via charge transfer. However, the optically excited electrons in $E_C$ tend to drift from the thin to thick areas because of the funneling effect for electron transfer along the thickness-varied $WS_2$ areas that feature a cascading band alignment. As a result, the excited electrons and generated $O^{2-}$ ions are prone to accumulate in the thick areas and lead to high reaction rates therein.

The high reaction selectivity can be also attributed to the exact positions of reactive reaction sites that are usually associated with the local density of dangling bonds or lattice vacancies and are located at flake edges. For clarity, we select five typical sites A−E distributed in different flake areas and compare the chemical reactivity among them. Owing to the presence of high densities of dangling bonds at the edges of each layer, the three peripheral sites A, C, and D tend to be more chemically active than the central sites B and E, resulting in an edge-initiated reaction spreading inwards (Movie S1, Supporting Information). In combination of the two factors above, the reaction rates at different sites can be categorized as $r_D > r_C >> r_A >> r_B$ and $r_D >> r_E$. The variation of chemical reactivity at different sites accounts well for the photodelamination behavior, which selectively stripes the excessive overlying layers above the desired monolayer base.

As mentioned, a grand challenge in 2D semiconductor synthesis is the control on the non-uniform nucleation and over-grown local islands. A highly selective post-etching technique that enables precisely stripping the excessive overlying islands would benefit the thickness uniformity management for attaining highly uniform monolayer wafers for electronic purpose. Evidently, this photodelamination features a remarkable etching selectivity that can be readily exploited for addressing the challenge. In Figure 3c (also Figure S2, Supporting Information), we further verify the feasibility of this scheme. With appropriate light illumination, the overlying island (i.e., the $2^{nd}$ layer, shown as a black area in the PL image) above the monolayer base can be completely removed. As a result, a wafer with uniform thickness distribution over the entire region is attained (See corresponding PL images in Figure 3c and Figure S2d, Supporting Information). Also, we confirm the applicability of this scheme to $MoS_2$, another van der Waals semiconductor (Figure S3, Supporting Information), indicating the universality of this scheme. Hence, it represents a possible solution to address the challenge in uniformity management confronted in 2D electronics.

### 2.5 Lattice quality of photoetched 1L $WS_2$

From the point of view of engineering, the trait of non-invasivity represents a highly desired but challenging requirement for atomic-layer etching. We then focus on evaluating the etching invasiveness and lattice quality of the photoetched (thinned) 1L $WS_2$. To this end, we employed an aberration-corrected, atomically resolved STEM to examine the lattice quality of the underlying layer after photoetching by making statistics on the density of atomic vacancies over multiple regions.[29]-[31] After transferring a partially photoetched and well rinsed flake onto a STEM grid (inset of **Figure 4**a), we succeeded in locating the as-thinned 1L area by identifying the ragged etching boundary, as shown in Figure 4. Note that the natural 1L/2L boundaries through mechanical exfoliation tend to be atomically sharp at the edges. Selected-area diffraction imaging on the thinned 1L area reveals well-preserved lattice integrity because only spotted diffraction patterns (Figure 4b) are observed without exhibiting any sign of circular patterns as exhibited by amorphous phases (e.g., inset of Figure 2a).

The high lattice quality of the thinned 1L can be also justified through counting the density of lattice vacancies. Figure 4c displays typical atomic images for three independent areas taken under annular dark-field mode. In such an imaging mode, the heavy atoms would exhibit higher brightness than light ones and, accordingly, the presence of atomic vacancies can be readily discerned from the reduction in brightness. In this way, the local regions containing sulfur vacancies can be clearly visualized and counted for statistical analysis. In Figure 4c, the sites of individual sulfur vacancies are all identified and marked by red arrows. A more detailed method to identify sulfur vacancies can be found in Figure S4 (Supporting Information). For statistical purposes, more than 20 independent regions with an area of $4 \times 4$ nm$^2$ were recorded (Figure S5, Supporting Information) for both the thinned and pristine 1L areas. For comparison, Figure 4d plots the statistical data on the densities of lattice vacancies, [V], for the two types of areas, where the values of average (A) and standard deviation (D) from Gaussian fittings were obtained. The average [V] level in the thinned 1L area is estimated to be $4.2\times10^{13}$ cm$^{-2}$, which is quite comparable to that in the pristine 1L area, suggesting insignificant etching invasivity to the preserved layers in the mild photodelamination process.

## 2.6 Electronic performance of thinned 1L $WS_2$

It is widely accepted that the electronic performance (e.g. charge mobility, $\mu$) of 2D semiconductors assessed through charge transport represents one of the most strict probes for material quality, because it is sensitive to both atomic defects and surfacial absorbates.[18,19] In addition to atomic defects, we also examine the possible adverse impact on the electronic transport of thinned 1L $WS_2$ from surfacial residues due to incomplete stripping of reaction products. For this purpose, we prepared field-effect transistors by placing the thinned 1L $WS_2$ on a clean BN dielectric and used it as conduction channel. **Figure 5**a shows the cryogenic four-terminal conductivity ($G$) versus gate voltage ($V_g$) for the transistor. In all measured $T$s, the device exhibits high switching ratios spanning 5 orders in magnitude in response to electric gating, indicating the well-preserved semiconducting nature of the thinned 1L flake.

Figure 5b summarizes the $\mu$ values (blue dots) in the entire $T$ regime, where a band-like transport behavior is seen, confirming the high lattice quality of the thinned 1L channel. Importantly, $\mu$ reaches 80 and 210 cm$^2$V$^{-1}$s$^{-1}$ at 293 and 10 K, respectively. Both values are comparable to those of mechanically exfoliated pristine (red squares) or the state-of-art synthesized 1L counterparts,[5,6] further proving the highly preserved lattice quality from the parent flakes. More attractively, $\mu$ of the thinned 1L $WS_2$ (~0.7 nm) can reach the level of 2.5-nm-thick ultrathin Si (silicon on insulator, SOI) FET,[32] revealing the unique advantages in surface conditions and thickness over silicon to overcome the challenges of the power-law performance degradation and the short-channel effect for More-Moore electronics.[1,3]

The merit of non-invasivity makes photodelamination a superior route over all the conventional methods.[8–17] In Figure 5c we compare room-$T$ $\mu$ of our thinned samples with those defined by other high-power methods.[9,15,20,33] In general, $\mu$ of our 1L sample is 1–3 orders higher in magnitude than the other methods and comparable to the mechanically exfoliated $WS_2$ samples[34,35] or ultrathin silicon.[32,36] Together with the spectral and crystallographic characterization, the high electronic performance represents consistent evidence that justifies the high crystalline quality in the thinned 1L flakes, which also proves the feasibility of exploiting the photodelamination as a non-invasive etching strategy for etching van der Waals semiconductors for high-performance electronics.[1–3]

## 3. Conclusion

We have observed an intriguing photodelamination phenomenon in van der Waals semiconductors and exploited its potential as a feasible atomic-layer etching strategy.

Micro-zone compositional analyses reveal the surfacial photooxidation reaction mechanism behind the photodelamination phenomenon. By using extensive spectral (PL and Raman), crystallographic (top-view and cross-sectional STEM, XPS and EDS), and electronic ($\mu$-$T$) characterization techniques, we consistently verified the remarkable merits of this mild etching strategy, including non-invasivity due to low excitation power, high etching selectivity, atomic etching precision, and processing convenience over previous methods. The results open a non-invasive atomic-layer etching route for thickness uniformity optimization in two-dimensional semiconductors toward high-performance atomic electronics.

## 4. Experimental Section

**Preparation of samples**. All $WS_2$ flakes were mechanically exfoliated onto viscoelastic PDMS substrates from synthesized crystals with Scotch tapes. The thickness information of local $WS_2$ areas was checked by PL and Raman spectra. A 30% hydrogen peroxide $H_2O_2$ solution was employed as a cleaner to remove the chemical groups bonded to the dangling bonds at the flake edges to activate the reaction sites. The liquid cleaner was directly dropped onto $WS_2$ flakes and the soaking time in oxidant drops was used as an index to evaluate the activation efficiency. We also note that the step of $H_2O_2$ pretreatment has a side effect; it can introduce lattice vacancies at a rate of $8.5\times10^{12}$ $cm^{-2}min^{-1}$ (Figure S6d, Supporting Information) into the lattice base (e.g., site B in Figure 3a). Long-time $H_2O_2$ pretreatment may lead to unintentional etching of the monolayer base during light illumination (Figure S7, Supporting Information). Thus, the duration of $H_2O_2$ exposure should be minimized to avoid damages to the lattice base.

**Simulation of temperature distribution.** To check the photothermal effect and investigate the local temperature of the samples under the illumination of a power density of 60 $nW/\mu m^2$, we simulated the steady-state temperature of the $WS_2$ flakes through a commercial software. The thermal conductivities are adopted as 116.8 and 0.16 W/(K·m) for $WS_2$ and PDMS substrate, respectively. The net optical absorption of 1L $WS_2$ is assumed to be 5%. The temperature of the sample rises to 0.08˚C at steady-state. To verify the accuracy of this method, we also simulated the temperature under the laser with 90 $mW/\mu m^2$, and the results are basically consistent with the previous literature.[10]

**Preparation of cross-sectional STEM sample**. The cross-sectional STEM specimens were fabricated by a lift-out method using focused ion beam (FIB) technique (FEI Helios 600i dual-beam system). The partially photodelaminated $WS_2$ flakes containing thinned 1L and pristine 2L areas were first transferred on $SiO_2$/Si substrates (Figure S9a, Supporting Information). Before FIB milling, amorphous carbon and metal Pt were deposited as protection layers onto the $WS_2/SiO_2$/Si stacks to avoid chemical contamination during milling. Figure S9b (Supporting Information) shows a typical cross-sectional slice for the overall quintuple Pt/carbon/$WS_2/SiO_2$/Si stacks. A typical atomically resolved cross-sectional image is given in Figure S9c (Supporting Information), where an etching step that separates the thinned 1L and pristine 2L areas can be clearly seen.

High-resolution ADF-STEM images were taken on a double aberration-corrected FEI Titan Cubed G2 60-300 S/TEM at 60 kV. To enhance the contrast of the sulfur sublattices, a medium-range ADF mode was used by adjusting the camera length properly. The probe current is set at 56 pA and the integration time is 8 s for collecting an image.

**Device fabrication and characterization.** The pristine and as-thinned 1L $WS_2$ samples were first transferred onto BN/$SiO_2$/Si substrates and etched by mixed $O_2$ and $CF_4$ plasma to define the transistor channels. Afterwards, an electron beam lithography system was employed to pattern source and drain electrodes, followed by thermal evaporation of 10 nm Ni/50 nm Au and standard lift-off. The devices were then mounted into the vacuum chamber of a probe station (CRX-6.5K, Lake Shore). Electrical measurements were performed with Keithley 2636B sourcemeters in a vacuum of $10^{-5}$ torr. The field-effect mobilities of the devices were calculated by using the equation $\mu = \frac{L}{W} \cdot \frac{1}{C_{\mathrm{ox}}} \cdot \frac{\partial G}{\partial V_g}$, where the $L/W$ is the aspect ratio of the device channel, $C_{\mathrm{ox}}$ is the capacitance of the BN/$SiO_2$ bilayer dielectrics.

## Supporting Information

Supporting Information is available from the Wiley Online Library or from the author.

## Acknowledgements

N. Xu and X. Pei contributed equally to this work. This work was supported by the National Key R&D Program of China (2021YFA1202903) and the National Natural Science Foundation of China (92264202, 61974060 and 61674080).

Received: ((will be filled in by the editorial staff))
Revised: ((will be filled in by the editorial staff))
Published online: ((will be filled in by the editorial staff))

[1] Y. Liu, X. Duan, H.-J. Shin, S. Park, Y. Huang, X. Duan, *Nature* **2021**, *591*, 43.

[2] F. Wu, H. Tian, Y. Shen, Z. Hou, J. Ren, G. Gou, Y. Sun, Y. Yang, T.-L. Ren, *Nature* **2022**, *603*, 259.

[3] S.-L. Li, K. Tsukagoshi, E. Orgiu, P. Samorì, *Chem. Soc. Rev.* **2016**, *45*, 118.

[4] J. Wang, X. Xu, T. Cheng, L. Gu, R. Qiao, Z. Liang, D. Ding, H. Hong, P. Zheng, Z. Zhang, Z. Zhang, S. Zhang, G. Cui, C. Chang, C. Huang, J. Qi, J. Liang, C. Liu, Y. Zuo, G. Xue, X. Fang, J. Tian, M. Wu, Y. Guo, Z. Yao, Q. Jiao, L. Liu, P. Gao, Q. Li, R. Yang, G. Zhang, Z. Tang, D. Yu, E. Wang, J. Lu, Y. Zhao, S. Wu, F. Ding, K. Liu, *Nat. Nanotechnol.* **2022**, *17*, 33.

[5] T. Li, W. Guo, L. Ma, W. Li, Z. Yu, Z. Han, S. Gao, L. Liu, D. Fan, Z. Wang, Y. Yang, W. Lin, Z. Luo, X. Chen, N. Dai, X. Tu, D. Pan, Y. Yao, P. Wang, Y. Nie, J. Wang, Y. Shi, X. Wang, *Nat. Nanotechnol.* **2021**, *16*, 1201.

[6] Q. Wang, J. Tang, X. Li, J. Tian, J. Liang, N. Li, D. Ji, L. Xian, Y. Guo, L. Li, Q. Zhang, Y. Chu, Z. Wei, Y. Zhao, L. Du, H. Yu, X. Bai, L. Gu, K. Liu, W. Yang, R. Yang, D. Shi, G. Zhang, *Natl. Sci. Rev.* **2022**, *9*, nwac077.

[7] L. Liu, T. Li, L. Ma, W. Li, S. Gao, W. Sun, R. Dong, X. Zou, D. Fan, L. Shao, C. Gu, N. Dai, Z. Yu, X. Chen, X. Tu, Y. Nie, P. Wang, J. Wang, Y. Shi, X. Wang, *Nature* **2022**, *605*, 69.

[8] N. Balakrishnan, Z. R. Kudrynskyi, E. F. Smith, M. W. Fay, O. Makarovsky, Z. D. Kovalyuk, L. Eaves, P. H. Beton, A. Patanè, *2D Mater.* **2017**, *4*, 025043.

[9] A. Castellanos-Gomez, M. Barkelid, A. M. Goossens, V. E. Calado, H. S. J. van der Zant, G. A. Steele, *Nano Lett.* **2012**, *12*, 3187.

[10] L. Hu, X. Shan, Y. Wu, J. Zhao, X. Lu, *Sci. Rep.* **2017**, *7*, 15538.

[11] Y. Rho, J. Pei, L. Wang, Z. Su, M. Eliceiri, C. P. Grigoropoulos, *ACS Appl. Mater. Interfaces* **2019**, *11*, 39385.

[12] V. K. Nagareddy, T. J. Octon, N. J. Townsend, S. Russo, M. F. Craciun, C. D. Wright, *Adv. Funct. Mater.* **2018**, *28*, 1804434.

[13] S. Xiao, P. Xiao, X. Zhang, D. Yan, X. Gu, F. Qin, Z. Ni, Z. J. Han, K. O. Ostrikov, *Sci. Rep.* **2016**, *6*, 19945.

[14] Y. Liu, H. Nan, X. Wu, W. Pan, W. Wang, J. Bai, W. Zhao, L. Sun, X. Wang, Z. Ni, *ACS Nano* **2013**, *7*, 4202.

[15] K. S. Kim, K. H. Kim, Y. Nam, J. Jeon, S. Yim, E. Singh, J. Y. Lee, S. J. Lee, Y. S. Jung, G. Y. Yeom, D. W. Kim, *ACS Appl. Mater. Interfaces* **2017**, *9*, 11967.

[16] M. Yamamoto, S. Nakaharai, K. Ueno, K. Tsukagoshi, *Nano Lett.* **2016**, *16*, 2720.

[17] K. K. Amara, L. Chu, R. Kumar, M. Toh, G. Eda, *APL Mater.* **2014**, *2*, 092509.

[18] S.-L. Li, K. Wakabayashi, Y. Xu, S. Nakaharai, K. Komatsu, W.-W. Li, Y.-F. Lin, A. Aparecido-Ferreira, K. Tsukagoshi, *Nano Lett.* **2013**, *13*, 3546.

[19] S. Ju, B. Liang, J. Zhou, D. Pan, Y. Shi, S. Li, *Nano Lett.* **2022**, *22*, 6671.

[20] J. Zhou, C. Zhang, L. Shi, X. Chen, T. S. Kim, M. Gyeon, J. Chen, J. Wang, L. Yu, X. Wang, K. Kang, E. Orgiu, P. Samorì, K. Watanabe, T. Taniguchi, K. Tsukagoshi, P. Wang, Y. Shi, S. Li, *Nat. Commun.* **2022**, *13*, 1844.

[21] A. Favron, E. Gaufrès, F. Fossard, A.-L. Phaneuf-L'Heureux, N. Y.-W. Tang, P. L. Lévesque, A. Loiseau, R. Leonelli, S. Francoeur, R. Martel, *Nat. Mater.* **2015**, *14*, 826.

[22] J. C. Kotsakidis, Q. Zhang, A. L. Vazquez De Parga, M. Currie, K. Helmerson, D. K. Gaskill, M. S. Fuhrer, *Nano Lett.* **2019**, *19*, 5205.

[23] A. Kuc, N. Zibouche, T. Heine, *Phys. Rev. B* **2011**, *83*, 245213.

[24] K. F. Mak, C. Lee, J. Hone, J. Shan, T. F. Heinz, *Phys. Rev. Lett.* **2010**, *105*, 136805.

[25] X. Zhang, X.-F. Qiao, W. Shi, J.-B. Wu, D.-S. Jiang, P.-H. Tan, *Chem. Soc. Rev.* **2015**, *44*, 2757.

[26] X. Lu, R. Wang, L. Hao, F. Yang, W. Jiao, P. Peng, F. Yuan, W. Liu, *Phys. Chem. Chem. Phys.* **2016**, *18*, 31211.

[27] K. Kang, K. Godin, Y. D. Kim, S. Fu, W. Cha, J. Hone, E.-H. Yang, *Adv. Mater.* **2017**, *29*, 1603898.

[28] H. Zeng, G.-B. Liu, J. Dai, Y. Yan, B. Zhu, R. He, L. Xie, S. Xu, X. Chen, W. Yao, X. Cui, *Sci. Rep.* **2013**, *3*.

[29] K. Fujisawa, B. R. Carvalho, T. Zhang, N. Perea-López, Z. Lin, V. Carozo, S. L. L. M. Ramos, E. Kahn, A. Bolotsky, H. Liu, A. L. Elías, M. Terrones, *ACS Nano* **2021**, *15*, 9658.

[30] A. M. van der Zande, P. Y. Huang, D. A. Chenet, T. C. Berkelbach, Y. You, G.-H. Lee, T. F. Heinz, D. R. Reichman, D. A. Muller, J. C. Hone, *Nat. Mater.* **2013**, *12*, 554.

[31] N. Xu, D. Hong, X. Pei, J. Zhou, F. Wang, P. Wang, Y. Tian, Y. Shi, S. Li, *ACS Photonics* **2022**, *9*, 3404.

[32] K. Uchida, H. Watanabe, A. Kinoshita, J. Koga, T. Numata, S. Takagi, in *Digest. International Electron Devices Meeting*, San Francisco, USA, **2002** 47–50.

[33] K. Sunamura, T. R. Page, K. Yoshida, T.-A. Yano, Y. Hayamizu, *J. Mater. Chem. C* **2016**, *4*, 3268.

[34] Y. Wang, T. Sohier, K. Watanabe, T. Taniguchi, M. J. Verstraete, E. Tutuc, *Appl. Phys. Lett.* **2021**, *118*, 102105.

[35] M. W. Iqbal, M. Z. Iqbal, M. F. Khan, M. A. Shehzad, Y. Seo, J. H. Park, C. Hwang, J. Eom, *Sci. Rep.* **2015**, *5*, 10699.

[36] M. Schmidt, M. C. Lemme, H. D. B. Gottlob, F. Driussi, L. Selmi, H. Kurz, *Solid-State Electron.* **2009**, *53*, 1246.

**Figure captions**

**Figure 1.** Photodelamination phenomenon in van der Waals semiconductors under low illumination fluencies. a) Schematic illustration of photodelamination phenomenon. The inset shows the comparison of spatial temperature distributions between photodelamination (60 nW/μm$^2$) and photothermal thinning (90 mW/μm$^2$). b) and c) Contrastive optical images, corresponding schematic diagrams and PL images (insets) before and after photodelamination. d) Raman spectra for the pristine 2L and photothinned 1L $WS_2$ flakes. e) PL spectra for the pristine 1L, 2L, and the photothinned 1L $WS_2$. f) Reaction rate ($r$) versus $1/T$. The slope of the dashed curve yields an activation energy $E_a$ of ~1.7±0.4 eV.

**Figure 2.** Compositional characterization and removal of reaction products. a) STEM image and corresponding elemental mappings for a fully reacted 1L $WS_2$ flake. b) XPS spectrum on the tungsten 4*f* core level from a partially reacted thick flake. c) and d) Surface morphologies of a typical delaminated area before and after rinse by deionized water. e) and f) Monolayered height difference near the reaction boundary verified by AFM and cross-sectional STEM.

**Figure 3.** Mechanism of reaction selectivity and possible application. a) Schematic diagram for the edge-initiated oxidation during photodelamination. b) Electron funneling transfer due to variation of bandgap with local thickness. c) Demonstration of its potential in stripping extra overlying islands on a synthesized $WS_2$ sample.

**Figure 4.** Crystallographic characterization for photoetched 1L $WS_2$. a) Typical STEM image near the reaction boundary, showing thinned 1L and pristine (unreacted) 2L areas. The inset shows the partially photothinned and well rinsed flake above a STEM grid. b) Electron diffraction pattern for the photothinned 1L area. c) Typical atomically resolved STEM images for three independent areas in the thinned 1L area from (a). Atomic sulfur vacancies are marked by red arrows. d) Comparison of statistical densities of atomic vacancies between thinned and pristine 1L areas.

**Figure 5.** Device characterization for photoetched 1L $WS_2$. a) Four-terminal channel conductivity ($G$) versus gate voltage ($V_g$) measured at cryogenic and room temperatures for a typical field-effect transistor consisting of thinned 1L $WS_2$ as conduction channel. b) Comparison of electron mobility ($\mu$) between thinned and pristinely exfoliated 1L $WS_2$ (inset: 2.5 nm Si) at different temperatures. All $\mu$ values of the thinned sample are comparable to the high-quality exfoliated $WS_2$ or ultrathin Si. c) Comparison of room-$T$ $\mu$ for as-thinned van der Waals semiconductors with those made by other methods (laser, electrochemical, plasma etching, and thermal diffusion) and ultrathin silicon with thickness ranging from 0.9 nm to 2.5 nm.

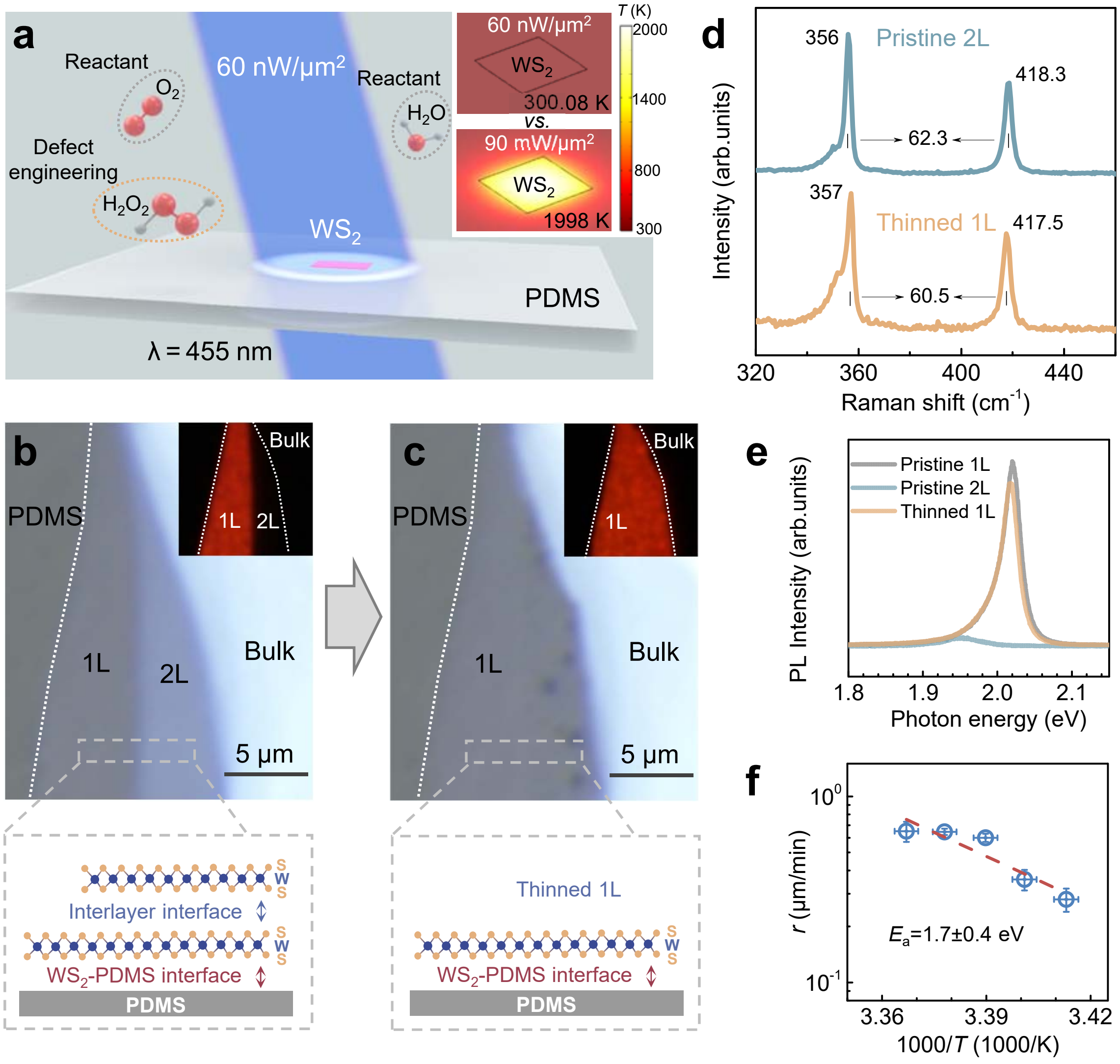

**Figure 1.**

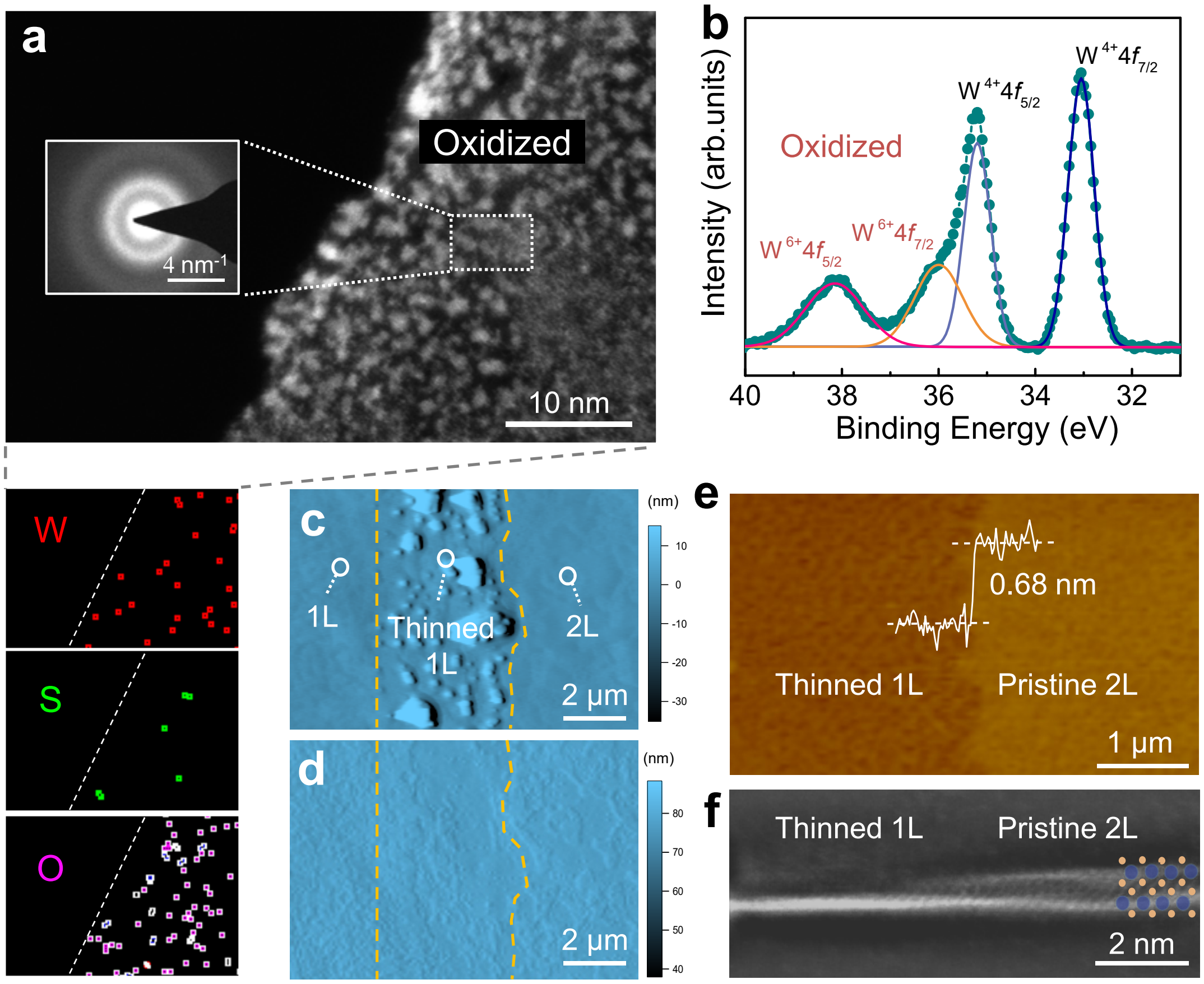

**Figure 2.**

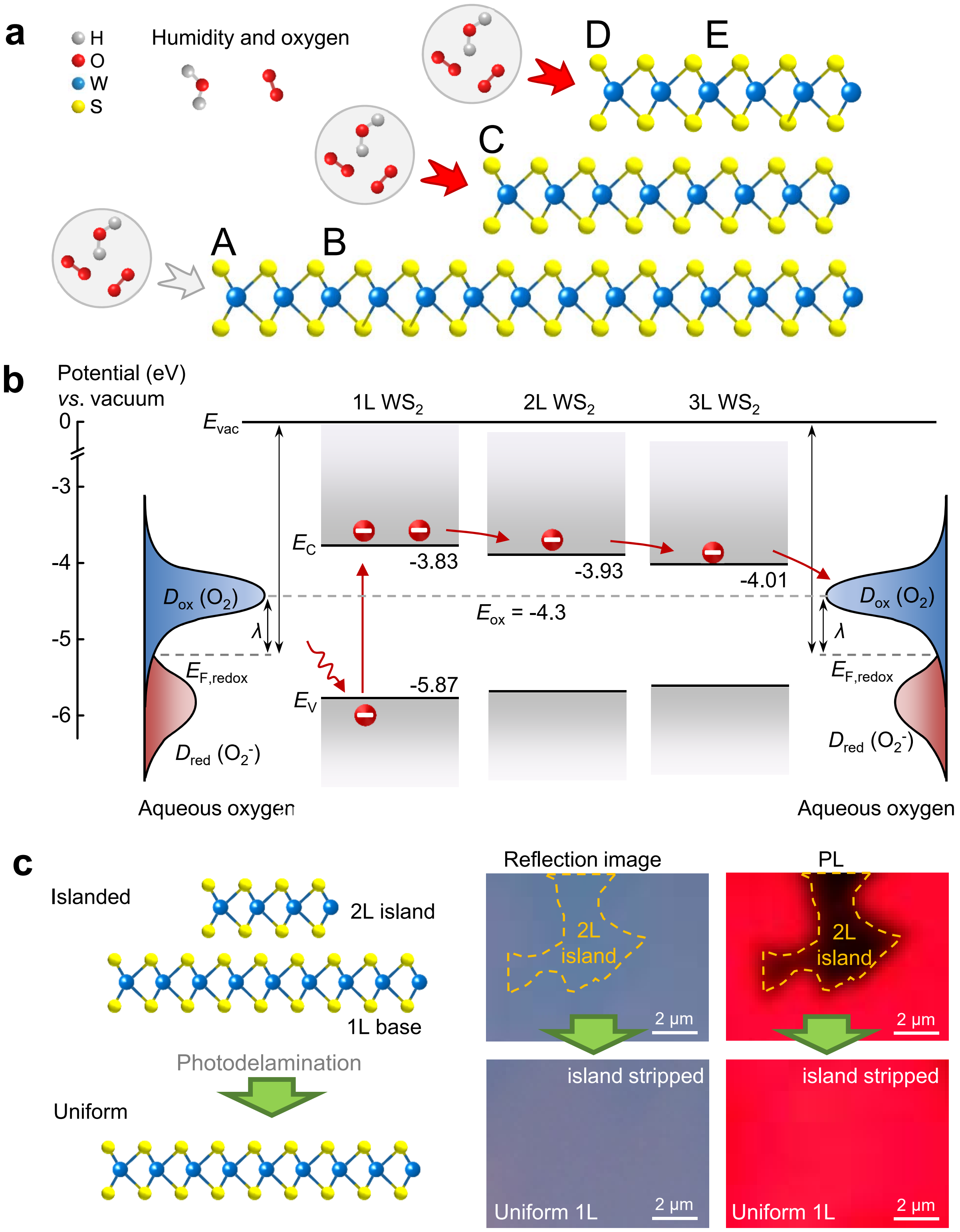

**Figure 3.**

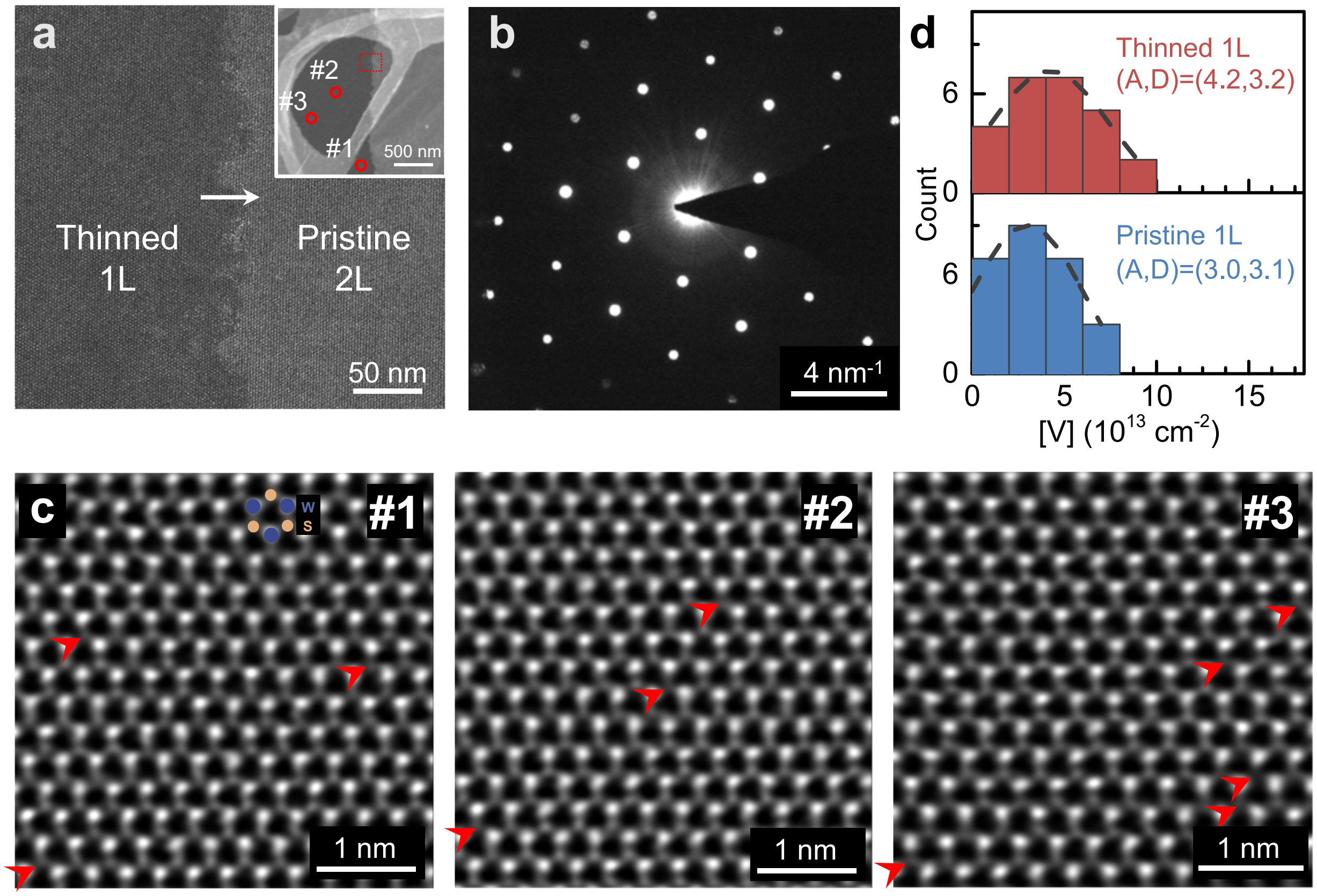

**Figure 4.**

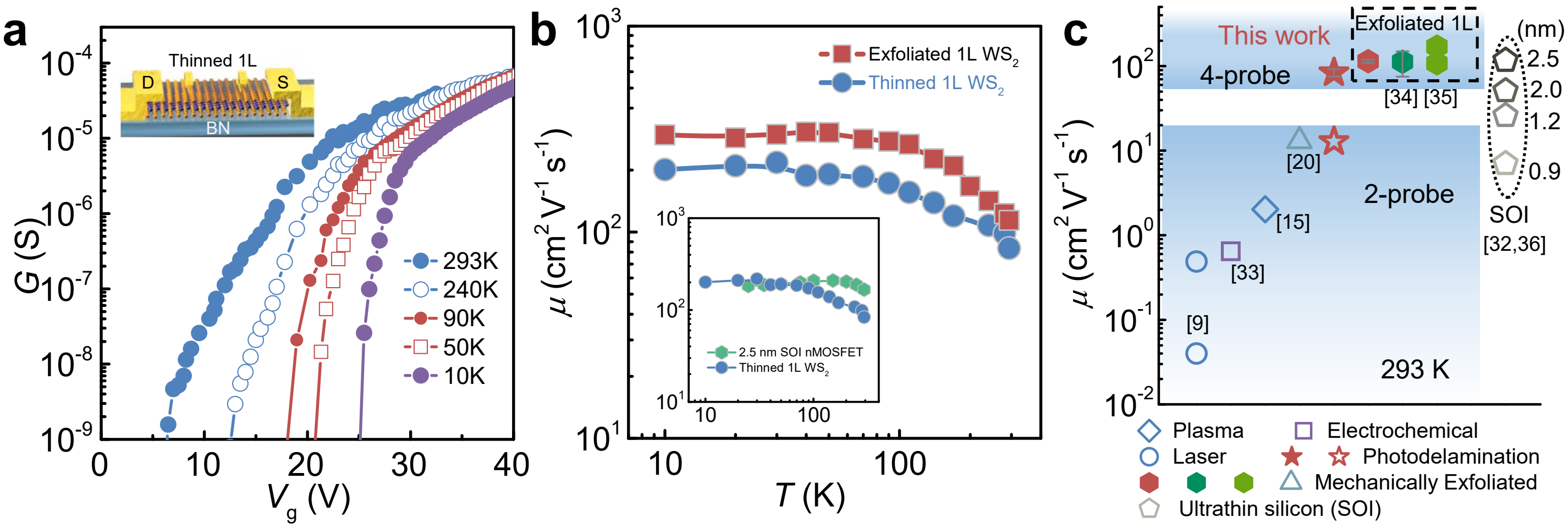

**Figure 5.**

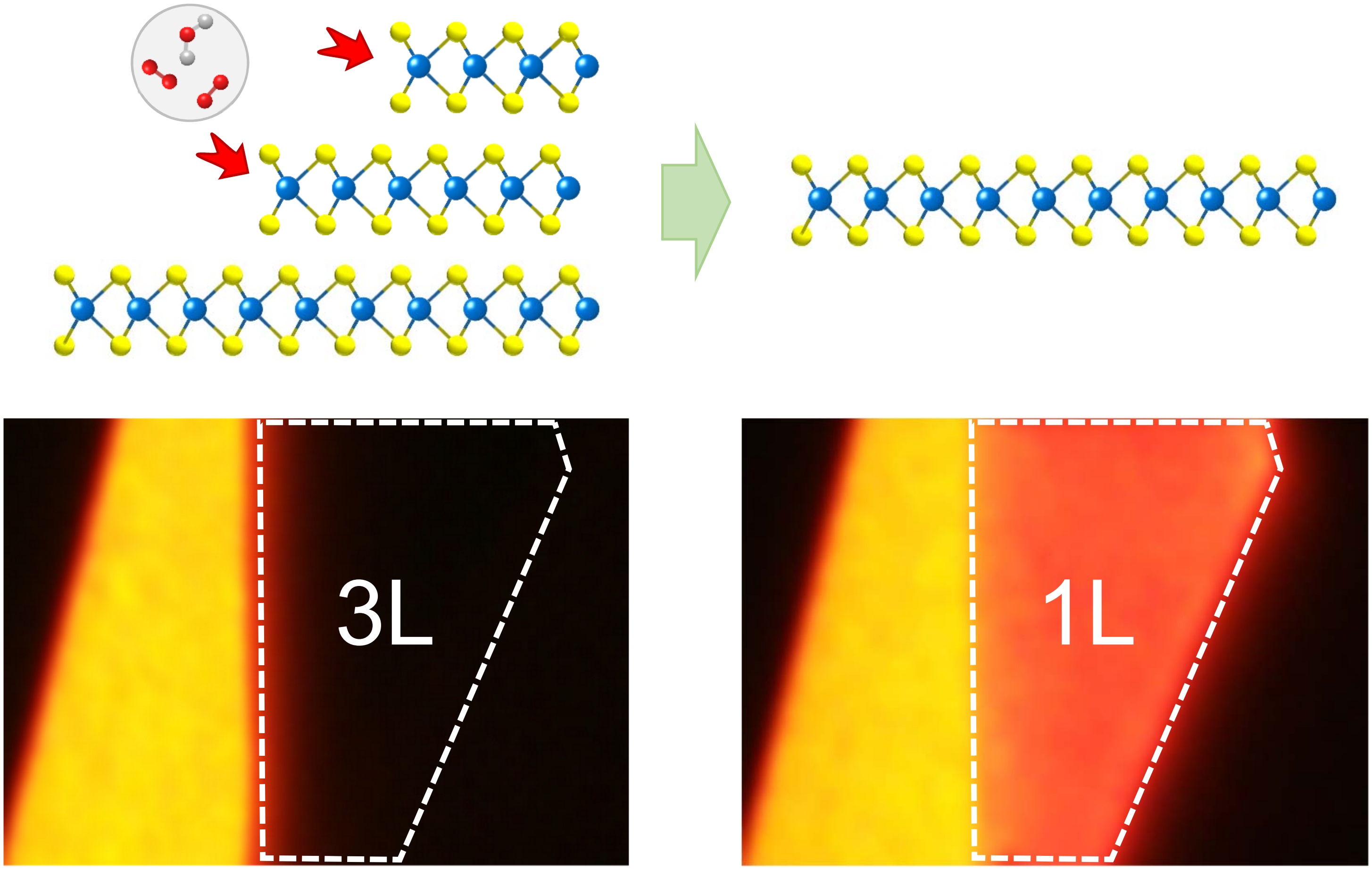

**TOC**

**Short text for ToC:** An intriguing photodelamination phenomenon that arises from ordered band alignment and resultant energy funneling effect is observed in nonuniform van der Waals semiconductors, which holds potential as a non-invasive but convenient atomic-layer etching strategy for thickness uniformity management in 2D van der Waals semiconductor wafers for electronic applications.

*Supporting Information*

# Non-Invasive Photodelamination of van der Waals Semiconductors for High-Performance Electronics

Ning Xu[1†], Xudong Pei[2†], Lipeng Qiu[1], Li Zhan[1], Peng Wang[3], Yi Shi[1*], Songlin Li[1*]

[1] National Laboratory of Solid-State Microstructures and Collaborative Innovation Center of Advanced Microstructures, School of Electronic Science and Engineering, Nanjing University, Nanjing 210023, China
[2] College of Engineering and Applied Sciences and Jiangsu Key Laboratory of Artificial Functional Materials, Nanjing University, Nanjing 210023, China
[3] Department of Physics, University of Warwick, CV4 7AL Coventry, UK

[†] These authors contributed equally to this work.
* Corresponding authors. Emails: sli@nju.edu.cn or yshi@nju.edu.cn.

## Contents

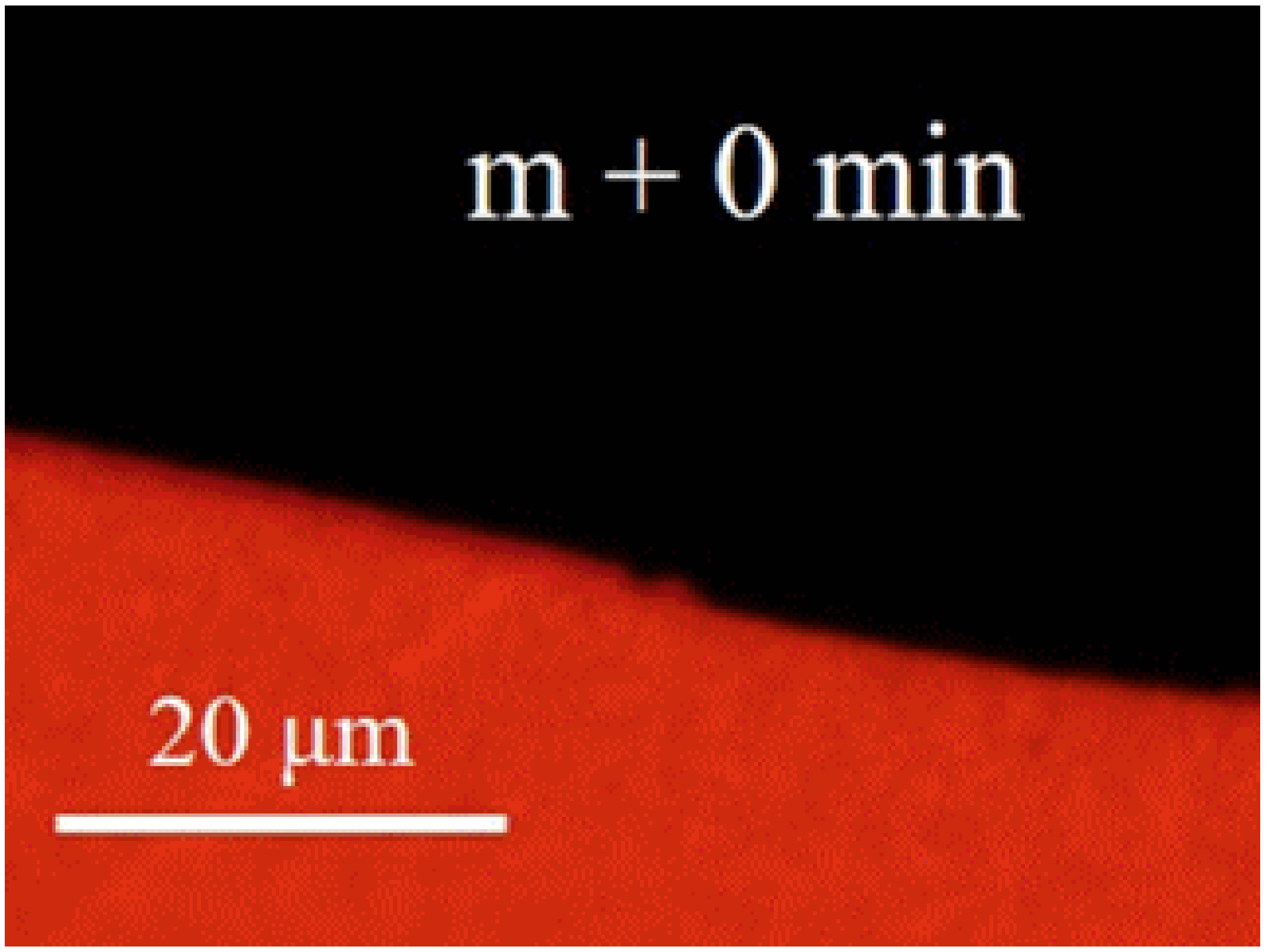

**MOVIE 1**

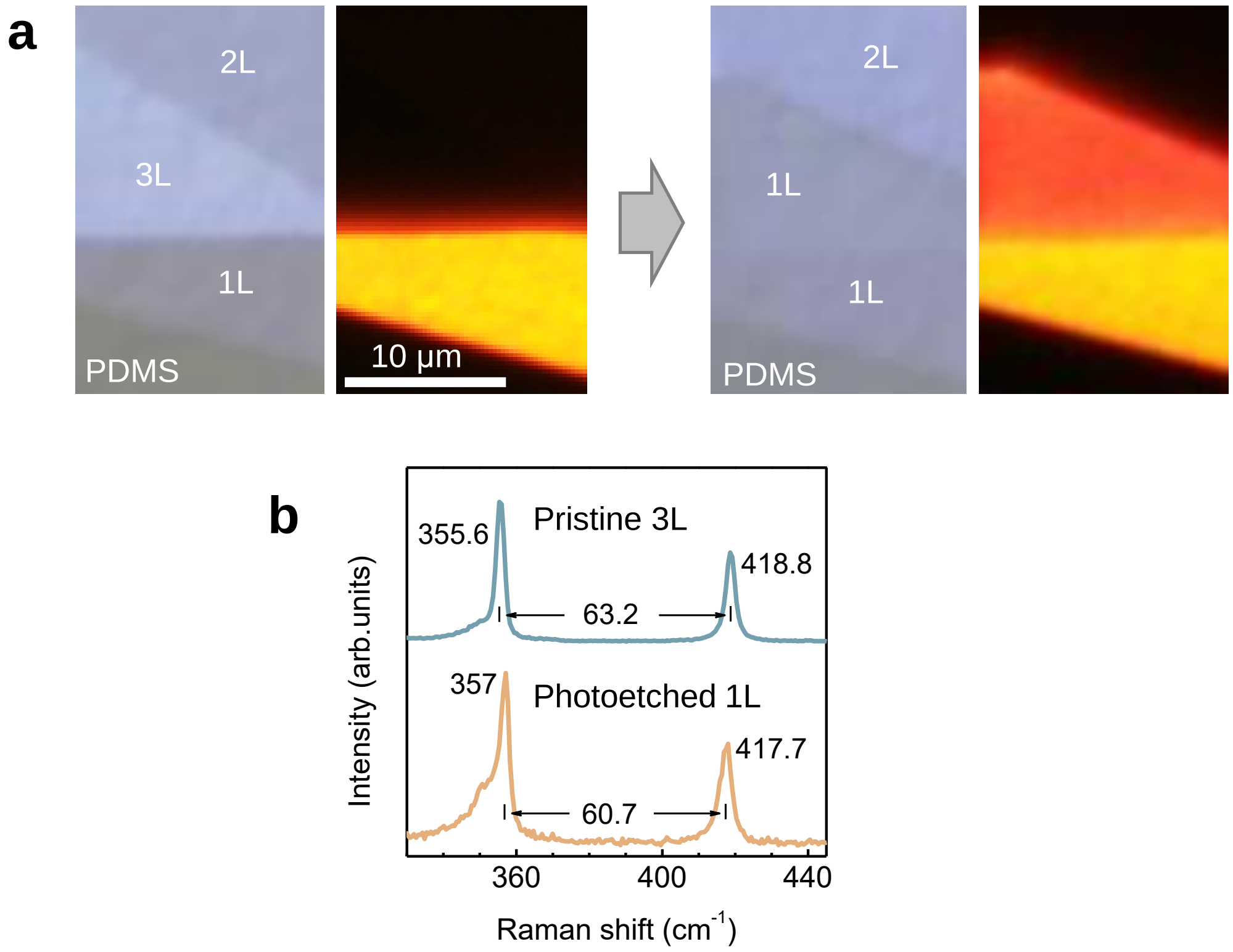

**Figure S1.** Nontrivial photodelamination behavior observed for a trilayer (3L) area. a) Contrastive optical images and corresponding PL images of a defect-engineered 3L $WS_2$ before and after photodelamination. The 3L area is directly converted into 1L. b) Corresponding Raman spectra for the 3L area before and after the nontrivial photodelamination .

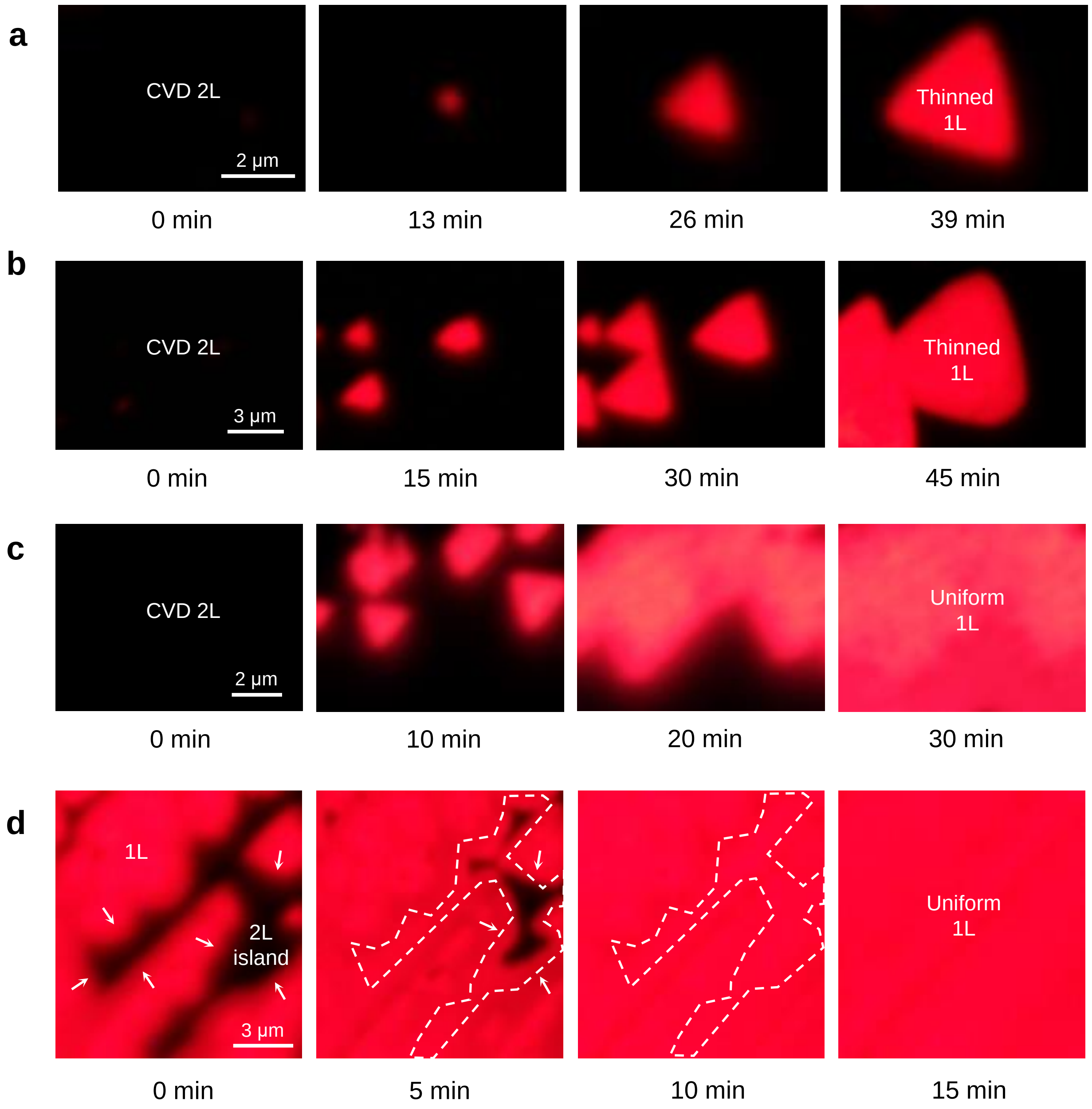

**Figure S2.** Local thickness tailoring in CVD synthesized $WS_2$ flakes on $SiO_2$/Si substrates. a) and b) Conversion from 2L to 1L from central areas via surfacial engineering. c) and d) Uniform management by stripping over-grown local islands above 1L bases.

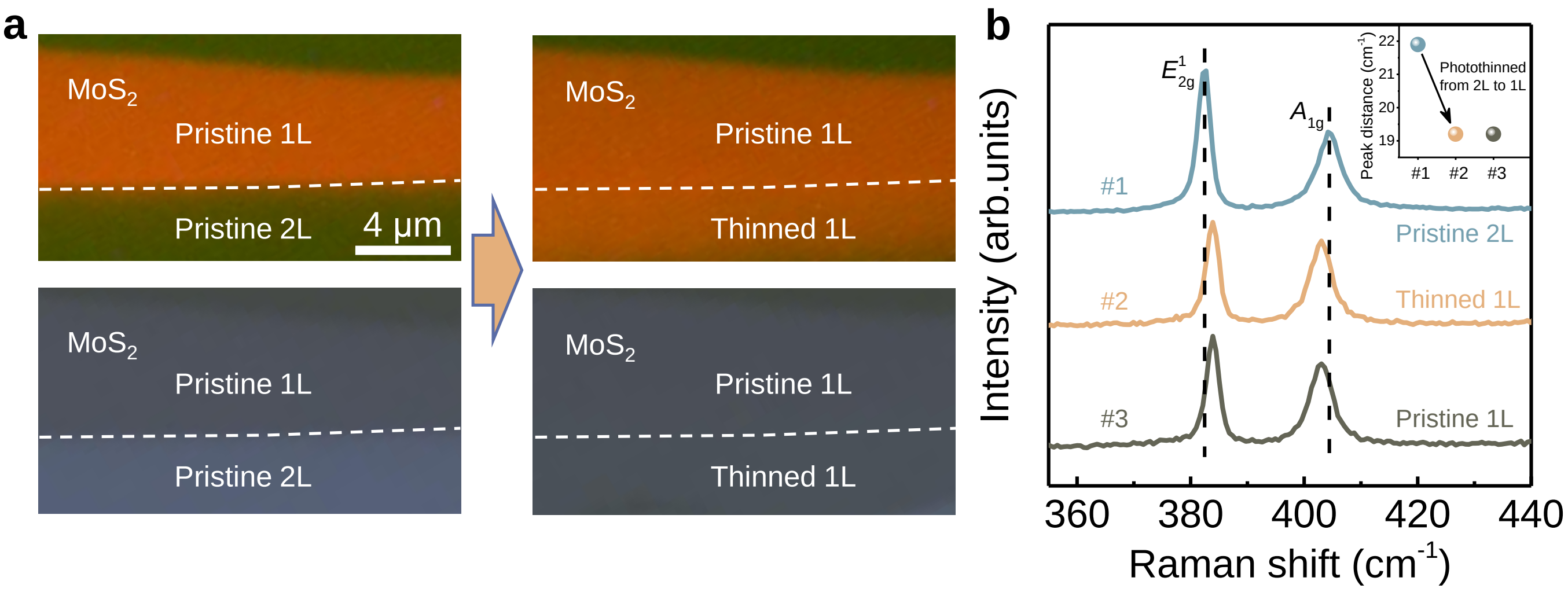

**Figure S3: Photodelamination in $MoS_2$.** a) Contrastive PL and reflection images for a $MoS_2$ flake with mixed 1L and 2L areas before and after photodelamination. **b**, Raman spectra for pristine 2L, thinned and pristine 1L $MoS_2$ areas. Inset: Variation of peak distance between the $E_{2g}{}^{1}$ and $A_{1g}$ modes after photodelamination.

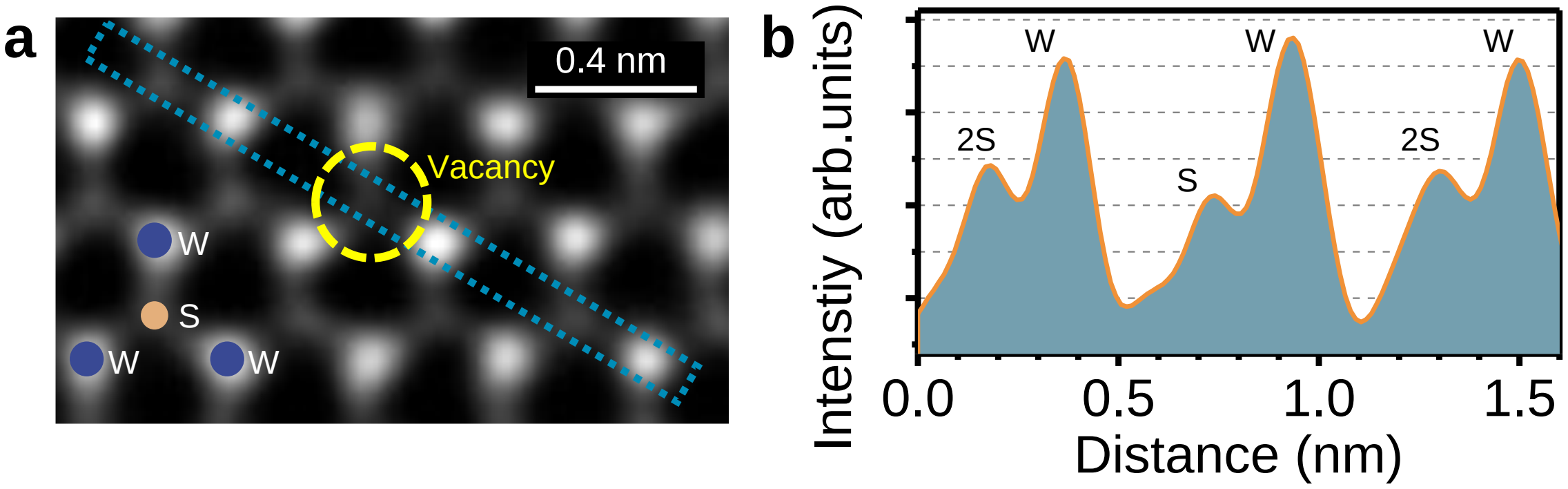

**Figure S4.** Methodology for identifying individual sulfur vacancies. a) Typical atomically resolved STEM image containing a sulfur vacancy (dashed yellow circle). b) Intensity profile along the atomic chains in the blue rectangle in (a). The reduced intensity at the sulfur site represents an individual atomic vacancy present in the $WS_2$ lattice.

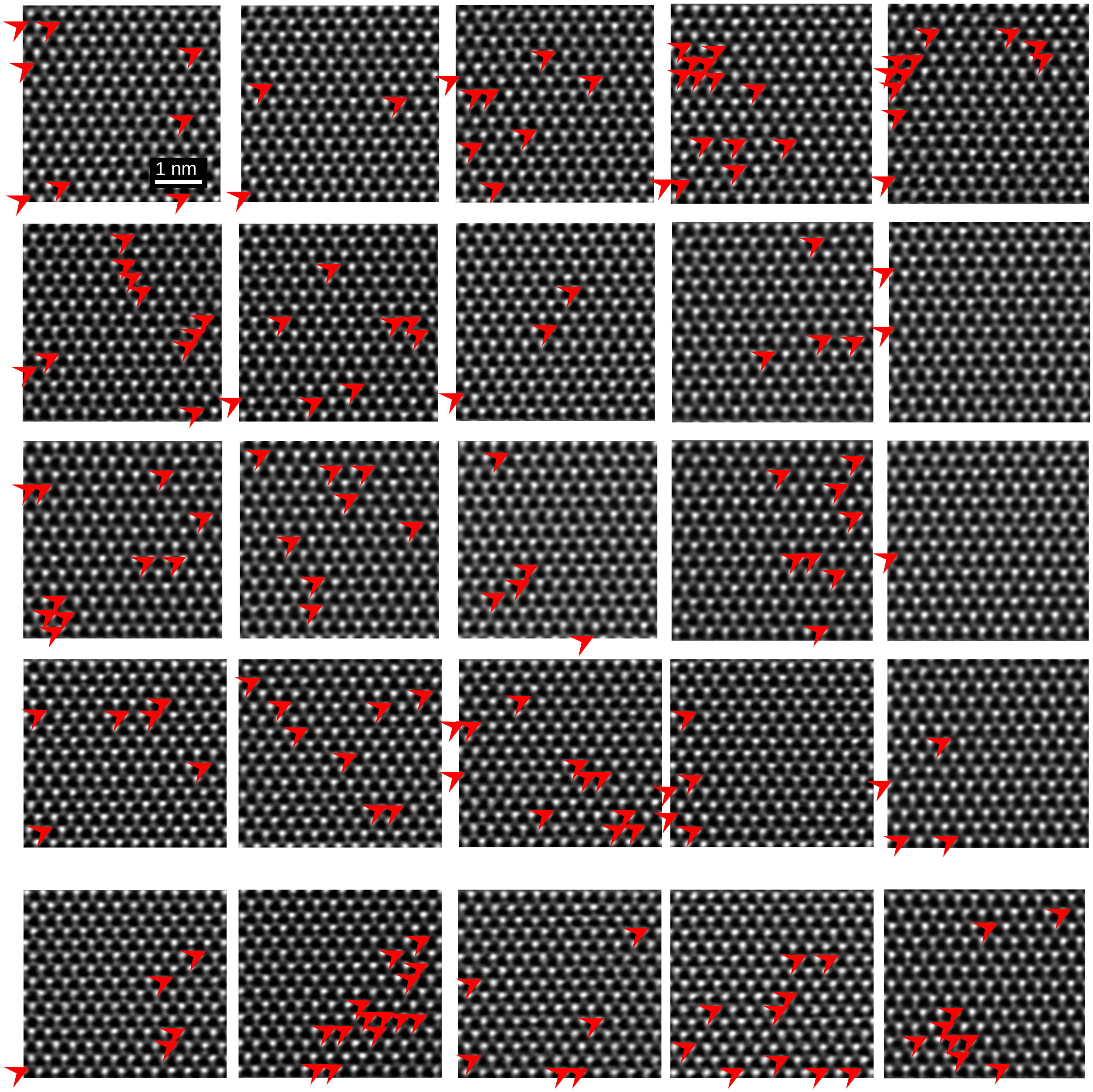

**Figure S5.** Atomically resolved STEM images for independent local areas from a photothinned 1L $WS_2$. The individual sulfur vacancies are labelled by red arrows.

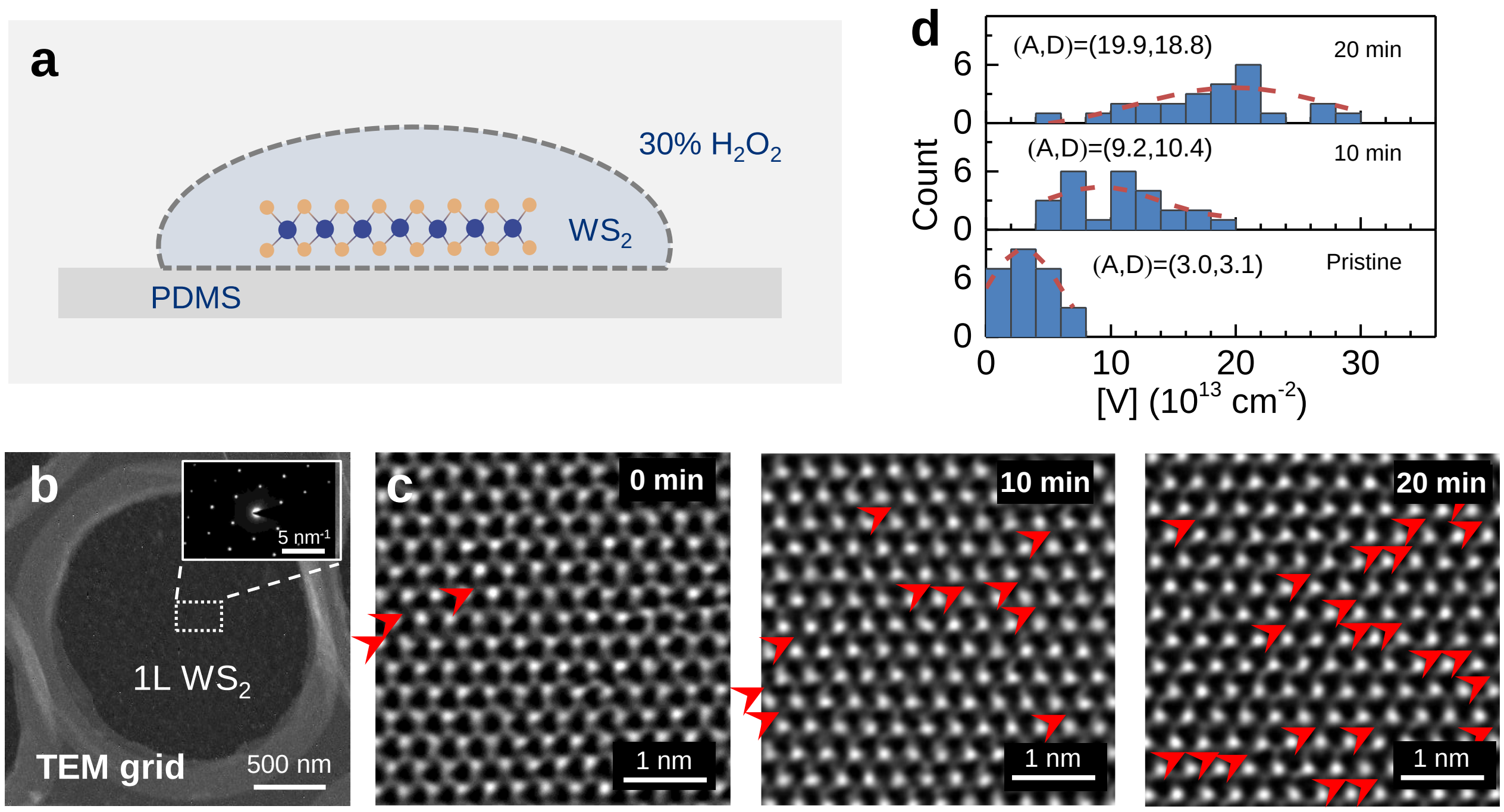

**Figure S6.** Edge activation in $H_2O_2$ solution and related side effect on lattice defects. a) Schematic illustration of $H_2O_2$ treatment on $WS_2$. b) Typical TEM image for a treated 1L $WS_2$ on a TEM grid. Inset: Electron diffraction pattern for near the rectangle area. The arrows denote the atomic sites with sulfur vacancies. c) Typical atomic images for $WS_2$ flakes treated for different durations from 0 to 20 min. d) Statistics on the density of atomic vacancies, [V], in samples pretreated for different durations. A and D denote the average and deviation values estimated from the Gaussian fittings.

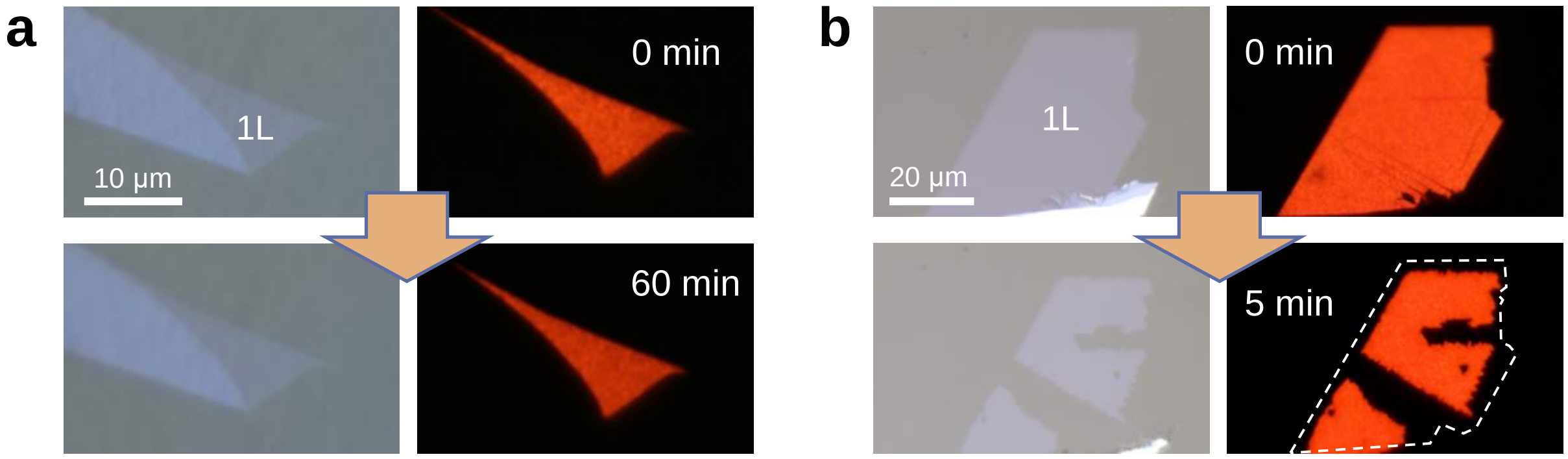

**Figure S7.** Acceleration effect of atomic vacancies on photooxidation in 1L $WS_2$. a) Contrastive PL images for a pristinely exfoliated 1L $WS_2$ before and after illumination for 60 min. No conspicuous photooxidation is seen. b) Contrastive PL images for a defect engineered 1L $WS_2$ (treated for 20 min with $H_2O_2$) before and after illumination for 5 min. Corrosion lengths up to 10 μm can be observed, indicating the remarkable acceleration effect of atomic vacancies on photooxidation.

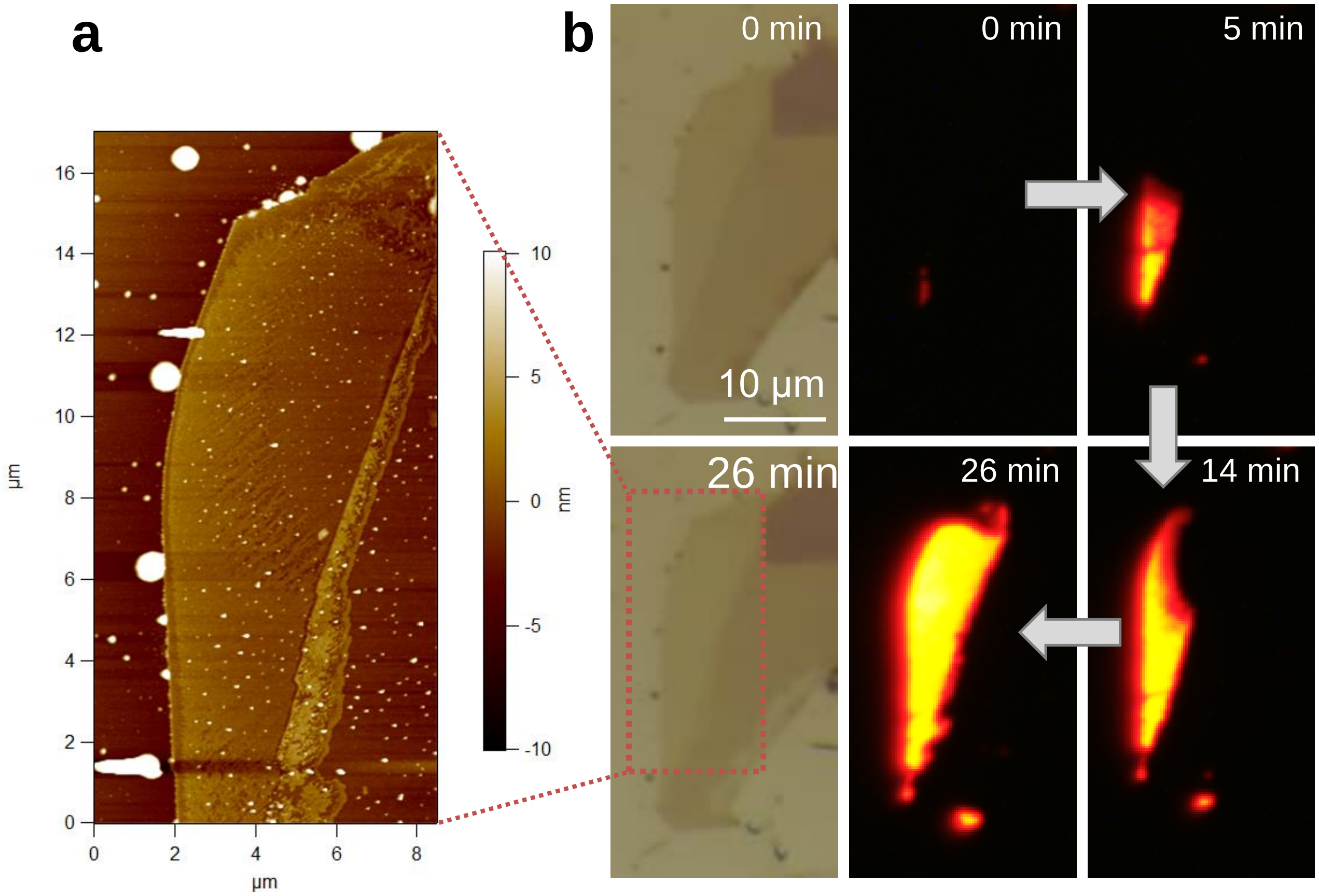

**Figure S8.** Alternative oxidant (TFSI, a Lewis acid) for activating the photodelamination behavior in $WS_2$ on $Al_2O_3$ substrates. a) AFM image of the photodelaminated area that is thinned into 1L. b) Serial optical and PL images tracking the photodelamination process at different stages.

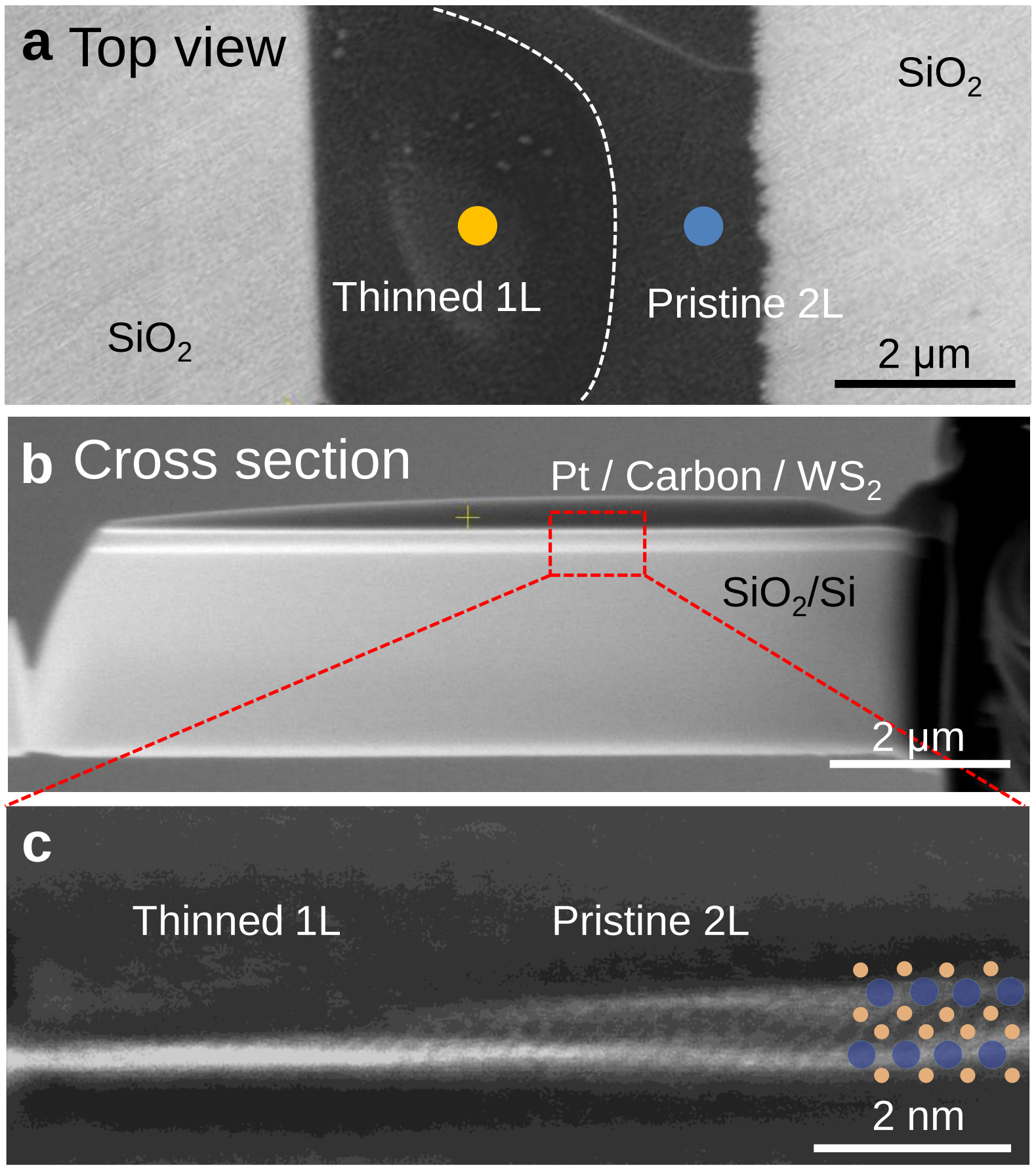

**Figure S9.** Preparation of STEM sample by focused ion beam milling for cross-sectional characterization. a) Top-view image to uncover the photoetched (thinned) and pristine areas, where the $WS_2$ flake is placed on a $SiO_2$ substrate. b) Cross-sectional view of the STEM sample just after FIB milling. c) Enlarged STEM image for identifying the photoetched (thinned) and pristine areas, where the etching boundary can be seen.